\documentclass[12pt,a4paper]{article}
\usepackage{fullpage}
\usepackage[centertags]{amsmath}
\allowdisplaybreaks[4]
\usepackage{amsfonts}
\usepackage{amssymb}
\usepackage{bm}
\usepackage{mathrsfs}
\usepackage{url}
\usepackage[dvips,hyperindex]{hyperref}
\begin{document}

\begin{flushright}
\setlength{\tabcolsep}{2pt}
\begin{tabular}{rl}
\textsf{arXiv:0805.0431}
\\
\textsf{4 May 2008}
\end{tabular}
\end{flushright}
\vspace{1cm}
\begin{center}
\Large\bfseries
Rates of Processes with Coherent Production of Different Particles
and the GSI Time Anomaly
\\[0.5cm]
\large\normalfont
Carlo Giunti
\\[0.5cm]
\normalsize\itshape
\setlength{\tabcolsep}{1pt}
\begin{tabular}{cl}
INFN, Sezione di Torino,
Via P. Giuria 1, I--10125 Torino, Italy
\end{tabular}
\end{center}
\begin{abstract}
With the help of an analogy with
a double-slit experiment,
it is shown that
the standard method of calculation of the rate
of an interaction process
by adding the rates of production
of all the allowed final states,
regardless of a possible coherence among them,
is correct.
It is a consequence of causality.
The claims
that the GSI time anomaly is due to the mixing of neutrinos
in the final state of the electron-capture process
are refuted.
It is shown that
the GSI time anomaly may be due to
quantum beats due to the existence of two
coherent energy levels of the decaying ion
with an extremely small energy splitting
(about $10^{-15}\,\text{eV}$)
and relative probabilities having a ratio of about 1/99.
\end{abstract}

The standard practice in the calculation of the rates
(cross sections and decay rates)
of interaction processes is to sum over the rates of production
of all the allowed channels with a defined number of particles in the final state,
regardless of a possible coherence among them.
This practice has been violated in recent papers
\cite{0801.1465,0801.2121,0801.3262}
claiming
that the anomalous oscillatory time modulation
of the electron-capture decays
\begin{equation}
{}^{140}\text{Pr}^{58+} \to {}^{140}\text{Ce}^{58+} + \nu_{e}
\,,
\qquad
{}^{142}\text{Pm}^{60+} \to ^{142}\text{Nd}^{60+} + \nu_{e}
\label{000}
\end{equation}
observed in a GSI experiment
\cite{0801.2079}
is due to neutrino mixing
(see
Refs.~\cite{Bilenky:1978nj,Bilenky:1987ty,hep-ph/9812360,hep-ph/0202058,hep-ph/0310238,hep-ph/0405172,hep-ph/0506083,hep-ph/0606054,Giunti-Kim-2007}).

In the simplest case of two-neutrino mixing,
the final $\nu_{e}$ in the processes in Eq.~(\ref{000})
is a coherent superposition of two massive neutrinos
conventionally called $\nu_{1}$ and $\nu_{2}$.
Neglecting the neutrino mass effects in the interaction
\cite{Giunti:1992cb,hep-ph/0102320,hep-ph/0306239,hep-ph/0402217,hep-ph/0608070,Giunti-Kim-2007},
the final electron neutrino state in the processes in Eq.~(\ref{000}) is
\begin{equation}
| \nu_{e} \rangle
=
\cos\vartheta \, | \nu_{1} \rangle
+
\sin\vartheta \, | \nu_{2} \rangle
\,,
\label{1331}
\end{equation}
where $\vartheta$ is the mixing angle.
In Refs.~\cite{0801.1465,0801.2121,0801.3262}
it is claimed that the interference of
the massive neutrinos in the final state
generates the observed time anomaly.
Such claim is in contradiction with standard calculations of decay rates,
in which the coherence of the final state is irrelevant:
the decay rates are calculated by adding the decay rates
with a massive neutrino in the final state.
The claim has been criticized in Refs.~\cite{0801.4639,0804.1099}.

Here I want to explain why the standard way of calculation of the rates
of interaction processes is correct.
For this purpose,
it is useful to clarify how interference occurs.

As an example, let us consider the
well-known
double-slit interference experiment with classical or quantum waves.
In a double slit experiment an
incoming plane wave packet hits a barrier with two tiny holes,
generating two outgoing spherical wave packets which propagate on the other side of the barrier.
The two outgoing waves are coherent,
since they are created with the same initial phase in the two holes.
Hence, the intensity after the barrier,
which is proportional to the squared modulus of the sum of the two outgoing waves,
exhibits interference effects.
The interference depends on the different path lengths of the two outgoing spherical waves
after the barrier.
Here, the important words for our discussion are ``after the barrier''.
The reason is that we can draw an analogy between the double-slit experiment
and an electron-capture decay process of the type in Eq.~(\ref{000}),
which can be schematically written as
\begin{equation}
\mathbb{I} \to \mathbb{F} + \nu_{e}
\,.
\label{003}
\end{equation}
Taking into account the neutrino mixing in Eq.~(\ref{1331}),
we have two different decay channels:
\begin{align}
\null & \null
\mathbb{I} \to \mathbb{F} + \nu_{1}
\,,
\label{004a}
\\
\null & \null
\mathbb{I} \to \mathbb{F} + \nu_{2}
\,.
\label{004b}
\end{align}
The initial state in the two decay channels
is the same.
In our analogy with the double-slit experiment,
the initial state $\mathbb{I}$ is analogous to
the incoming wave packet.
The two final states
$ \mathbb{F} + \nu_{1} $
and
$ \mathbb{F} + \nu_{2} $
are analogous to the two outgoing wave packets.
The different weights of $ \nu_{1} $ and $ \nu_2 $ production
due to a possible $ \theta \neq \pi/4 $
correspond to different sizes of the two holes in the barrier.

In the analogy,
the decay rate of $\mathbb{I}$ corresponds to the fraction of intensity of the incoming wave
which crosses the barrier.
I think that it is clear that the fraction of intensity of the incoming wave
depends only on the sizes of the holes.
It does not depend on the interference effect which occurs after the wave has passed through the barrier.
In a similar way,
the decay rate of $\mathbb{I}$ cannot depend on the interference of $\nu_{1}$ and $\nu_{2}$
which occurs after the decay has happened.

Of course,
neutrino oscillations caused by the interference of $\nu_{1}$ and $\nu_{2}$
can occur after the decay,
in analogy with the occurrence of interference of the outgoing waves in the double-slit experiment,
regardless of the fact that the decay rate is
the incoherent sum of the rates of production of $\nu_{1}$ and $\nu_{2}$
and the fraction of intensity of the incoming wave
which crosses the barrier is the incoherent sum of the fractions of intensity of the incoming wave which
pass trough the two holes.

The above argument
is a simple consequence of causality:
the interference of $\nu_{1}$ and $\nu_{2}$ occurring after the decay
cannot affect the decay rate.

Causality is explicitly violated in Ref.~\cite{0801.1465},
where the decaying ion is described by a wave packet,
but it is claimed that there is a selection of the momenta of the ion
caused by a final neutrino momentum splitting due to the mass difference
of $\nu_{1}$ and $\nu_{2}$.
This selection violates causality.
In the double-slit analogy,
the properties of the outgoing wave packets are determined by the properties
of the incoming wave packet,
not vice versa.
In a correct treatment, all the
momentum distribution of the wave packet of the ion contributes to the decay,
generating appropriate neutrino wave packets.

The authors of Refs.\cite{0801.2121,0801.3262}
use a different approach:
they calculate the decay rate with the final neutrino state
\begin{equation}
| \nu \rangle
=
\sum_{k} | \nu_{k} \rangle
\,.
\label{005}
\end{equation}
This state is different from the standard electron neutrino state,
which is given by
\begin{equation}
| \nu_{e} \rangle
=
\sum_{k} U_{ek}^* \, | \nu_{k} \rangle
\,,
\label{021}
\end{equation}
where $U$ is the mixing matrix
(in the two-neutrino mixing approximation of Eq.~(\ref{1331}),
$U_{e1}=\cos\vartheta$
and
$U_{e2}=\sin\vartheta$).

Omitting the complications of time-dependent perturbation theory used in Refs.\cite{0801.2121,0801.3262},
one can try to calculate the corresponding decay probability
in the framework of standard Quantum Field Theory:
\begin{equation}
P_{\mathbb{I} \to \mathbb{F} + \nu}
=
| \langle \nu, \mathbb{F} | \mathsf{S} | \mathbb{I} \rangle |^2
=
\left| \sum_{k} \langle \nu_{k}, \mathbb{F} | \mathsf{S} | \mathbb{I} \rangle \right|^2
\,,
\label{011}
\end{equation}
where $\mathsf{S}$ is the S-matrix operator.
The decay rate is obtained from the decay probability
by integrating over the phase space.

Considering the S-matrix operator at first order in perturbation theory,
\begin{equation}
\mathsf{S}
=
1
- i \int \text{d}^{4}x \, \mathscr{H}_{W}(x)
\,,
\label{012}
\end{equation}
with the effective four-fermion interaction Hamiltonian
\begin{align}
\mathscr{H}_{W}(x)
= 
\null & \null
\frac{G_F}{\sqrt{2}} \, \cos\theta_{\text{C}}
\,
\bar{\nu}_{e}(x) \gamma_{\rho} ( 1 - \gamma^5 ) e(x)
\,
\bar{n}(x) \gamma^{\rho} ( 1 - g_A \gamma^5 ) p(x)
\nonumber
\\
=
\null & \null
\frac{G_F}{\sqrt{2}} \, \cos\theta_{\text{C}}
\,
\sum_{k} \! U_{ek}^{*} \bar{\nu}_{k}(x) \gamma_{\rho} ( 1 - \gamma^5 ) e(x)
\,
\bar{n}(x) \gamma^{\rho} ( 1 - g_A \gamma^5 ) p(x)
\,,
\label{013}
\end{align}
where $\theta_{\text{C}}$ is the Cabibbo angle,
one can write the matrix elements in Eq.~(\ref{011}) as
\begin{equation}
\langle \nu_{k}, \mathbb{F} | \mathsf{S} | \mathbb{I} \rangle
=
U_{ek}^{*} \mathcal{M}_{k}
\,,
\label{014}
\end{equation}
with
\begin{equation}
\mathcal{M}_{k}
=
\frac{G_F}{\sqrt{2}} \, \cos\theta_{\text{C}}
\,
\langle \nu_{k}, \mathbb{F} |
\bar{\nu}_{k}(x) \gamma_{\rho} ( 1 - \gamma^5 ) e(x)
\,
\bar{n}(x) \gamma^{\rho} ( 1 - g_A \gamma^5 ) p(x)
| \mathbb{I} \rangle
\,.
\label{015}
\end{equation}
Therefore,
the decay probability is given by
\begin{equation}
P_{\mathbb{I} \to \mathbb{F} + \nu}
=
\left| \sum_{k} U_{ek}^{*} \mathcal{M}_{k} \right|^2
\,.
\label{016a}
\end{equation}
This decay probability is different from the standard one
\cite{Shrock:1980vy,McKellar:1980cn,Kobzarev:1980nk,Shrock:1981ct,Shrock:1981wq,hep-ph/0608070,Giunti-Kim-2007},
which is obtained by summing incoherently over the probabilities of decay
into the different massive neutrinos final states weighted by the
corresponding element of the mixing matrix:
\begin{equation}
P
=
\sum_{k} |U_{ek}|^2 \left| \mathcal{M}_{k} \right|^2
\,.
\label{016b}
\end{equation}

The analogy with the double-slit experiment
and the causality argument discussed above support the correctness of the standard
decay probability $P$.
Another argument against the decay probability $P_{\mathbb{I} \to \mathbb{F} + \nu}$
is that in the limit of massless neutrinos it does not
reduce to the decay probability in the Standard Model:
\begin{equation}
P_{\text{SM}}
=
\left| \mathcal{M}_{\text{SM}} \right|^2
\,,
\label{016}
\end{equation}
with
\begin{equation}
\mathcal{M}_{\text{SM}}
=
\frac{G_F}{\sqrt{2}} \, \cos\theta_{\text{C}}
\,
\langle \mathbb{F}, \nu_{e}^{\text{SM}} |
\bar{\nu}_{e}^{\text{SM}}(x) \gamma_{\rho} ( 1 - \gamma^5 ) e(x)
\,
\bar{n}(x) \gamma^{\rho} ( 1 - g_A \gamma^5 ) p(x)
| \mathbb{I} \rangle
\,,
\label{017}
\end{equation}
where $\nu_{e}^{\text{SM}}$ is the SM massless electron neutrino.
Indeed,
for the matrix elements $\mathcal{M}_{k}$ we have
\begin{equation}
\mathcal{M}_{k}
\xrightarrow[m_{k}\to0]{}
\mathcal{M}_{\text{SM}}
\,,
\label{018}
\end{equation}
leading to
\begin{equation}
P_{\mathbb{I} \to \mathbb{F} + \nu}
\xrightarrow[m_{k}\to0]{}
\left| \mathcal{M}_{\text{SM}} \right|^2
\left| \sum_{k} U_{ek}^{*} \right|^2
\,.
\label{019}
\end{equation}
This is different from the SM decay probability in Eq.~(\ref{016}).
Notice that the contribution of the elements
of the mixing matrix should disappear automatically
in the limit $m_{k}\to0$.
In fact, even in the SM one can define the three massless flavors neutrinos
$\nu_{e}$,
$\nu_{\mu}$,
$\nu_{\tau}$
as arbitrary unitary linear combinations of three massless neutrinos
$\nu_{1}$,
$\nu_{2}$,
$\nu_{3}$.
However,
all physical quantities are independent of such an arbitrary transformation.

We conclude that the state in Eq.~(\ref{005})
does not describe the neutrino emitted in an electron-capture decay process of the type in Eq.~(\ref{003}).
Notice that the state in Eq.~(\ref{005}) is not even properly normalized
to describe one particle ($\langle\nu|\nu\rangle=3$).

The correct normalized state
($\langle\nu_{e}|\nu_{e}\rangle=1$)
which describes
the electron neutrino emitted in an electron-capture decay processes of the type in Eq.~(\ref{003})
is
\begin{align}
| \nu_{e} \rangle
=
\null & \null
\left( \sum_{j}  | \langle \nu_{j}, \mathbb{F} | \mathsf{S} | \mathbb{I} \rangle |^2 \right)^{-1/2}
\sum_{k} | \nu_{k} \rangle \, \langle \nu_{k}, \mathbb{F} | \mathsf{S} | \mathbb{I} \rangle
\nonumber
\\
=
\null & \null
\left( \sum_{j}  | U_{ej} |^2 | \mathcal{M}_{j} |^2 \right)^{-1/2}
\sum_{k} U_{ek}^{*} \mathcal{M}_{k} \, | \nu_{k} \rangle
\,.
\label{121}
\end{align}
In experiments which are not sensitive to the differences of the neutrino masses,
as neutrino oscillation experiments,
we can approximate
$
\mathcal{M}_{k} \simeq \overline{\mathcal{M}}
$
and the state (\ref{121})
reduces to the standard electron neutrino state in Eq.~(\ref{021})
(apart for an irrelevant phase $\overline{\mathcal{M}}/|\overline{\mathcal{M}}|$).

With the electron neutrino state in Eq.~(\ref{121}),
the decay probability is given by
\begin{equation}
P_{\mathbb{I} \to \mathbb{F} + \nu_{e}}
=
| \langle \nu_e, \mathbb{F} | \mathsf{S} | \mathbb{I} \rangle |^2
=
\sum_{k} \left| \langle \nu_{k}, \mathbb{F} | \mathsf{S} | \mathbb{I} \rangle \right|^2
=
\sum_{k} |U_{ek}|^2 \left| \mathcal{M}_{k} \right|^2
\,.
\label{123}
\end{equation}
This is the correct standard result in Eq.~(\ref{016b}):
the decay probability is given by the incoherent sum over the probabilities of decay
into different massive neutrinos weighted by the
corresponding element of the mixing matrix.

Using Eq.~(\ref{018}) and the unitarity of the mixing matrix,
one can also easily check that $P_{\mathbb{I} \to \mathbb{F} + \nu_{e}}$
reduces to $P_{\text{SM}}$ in Eq.~(\ref{016})
in the massless neutrino limit.

Finally,
let me emphasize that,
although the GSI time anomaly cannot be due to effects of neutrino mixing
in the final state of the electron-capture process,
it can be due to interference effects in the initial state.
For example,
there could be an interference between two coherent energy states of the
decaying ion which produces quantum beats
(see, for example, Ref.~\cite{Carter-Huber-2000}).
Also in this case we can draw an analogy with a double-slit experiment.
However, we must change the setup,
considering a double-slit experiment with two coherent sources of incoming waves.
In this case,
the two incoming waves interfere at the holes in the barrier,
leading to a modulation of the intensity which crosses the barrier.
The role of causality is clear:
the interference effect is due to the different phases
of the two coherent incoming waves at the holes,
which have been developed during the propagation of the two waves
along different path lengths before reaching the barrier.
Analogously,
quantum beats in the GSI experiment can be due to
interference of two coherent energy states of the
decaying ion which develop different phases before the decay.
If the measuring apparatus which monitors the ions
with a frequency of
the order of the revolution frequency in the ESR storage ring, about 2 MHz,
does not distinguish
between the two states, their coherence is preserved for a long time.

If the two energy states of the decaying ion
$\mathbb{I}_{1}$ and $\mathbb{I}_{2}$
are produced at the time $t=0$ with amplitudes
$\mathcal{A}_{1}$ and $\mathcal{A}_{2}$
(with $ |\mathcal{A}_{1}|^2 + |\mathcal{A}_{2}|^2 = 1 $),
we have
\begin{equation}
| \mathbb{I}(t=0) \rangle
=
\mathcal{A}_{1} \, | \mathbb{I}_{1} \rangle
+
\mathcal{A}_{2} \, | \mathbb{I}_{2} \rangle
\,.
\label{201}
\end{equation}
Assuming, for simplicity, that the two states
with energies
$E_{1}$ and $E_{2}$
have the same decay rate $\Gamma$,
at the time $t$
we have
\begin{equation}
| \mathbb{I}(t) \rangle
=
\left(
\mathcal{A}_{1} \, e^{ - i E_{1} t } \, | \mathbb{I}_{1} \rangle
+
\mathcal{A}_{2} \, e^{ - i E_{2} t } \, | \mathbb{I}_{2} \rangle
\right)
e^{ - \Gamma t / 2 }
\,.
\label{202}
\end{equation}
The probability of electron capture at the time $t$ is given by
\begin{equation}
P_{\text{EC}}(t)
=
| \langle \nu_{e}, \mathbb{F} | \mathsf{S} | \mathbb{I}(t) \rangle |^2
=
\left[
1
+
A \, \cos\!\left( \Delta{E} t + \varphi \right)
\right]
\overline{P}_{\text{EC}}
\,
e^{ - \Gamma t }
\,,
\label{203}
\end{equation}
where
$ A \equiv 2 |\mathcal{A}_{1}| |\mathcal{A}_{2}| $,
$ \Delta{E} \equiv E_{2} - E_{1} $,
\begin{equation}
\overline{P}_{\text{EC}}
=
| \langle \nu_{e}, \mathbb{F} | \mathsf{S} | \mathbb{I}_{1} \rangle |^2
=
| \langle \nu_{e}, \mathbb{F} | \mathsf{S} | \mathbb{I}_{2} \rangle |^2
\,,
\label{204}
\end{equation}
and $\varphi$ is a constant phase which takes into account possible phase differences of
$\mathcal{A}_{1}$ and $\mathcal{A}_{2}$
and of
$ \langle \nu_{e}, \mathbb{F} | \mathsf{S} | \mathbb{I}_{1} \rangle $
and
$ \langle \nu_{e}, \mathbb{F} | \mathsf{S} | \mathbb{I}_{2} \rangle $.

The fit of GSI data presented in Ref.~\cite{0801.2079} gave
\begin{align}
\null & \null
\Delta{E}({}^{140}\text{Pr}^{58+})
=
( 5.86 \pm 0.07 ) \times 10^{-16} \, \text{eV}
\,,
\null & \null
\quad
\null & \null
A({}^{140}\text{Pr}^{58+})
=
0.18 \pm 0.03
\,,
\label{211}
\\
\null & \null
\Delta{E}({}^{142}\text{Pm}^{60+})
=
( 5.82 \pm 0.18 ) \times 10^{-16} \, \text{eV}
\,,
\null & \null
\quad
\null & \null
A({}^{142}\text{Pm}^{60+})
=
0.23 \pm 0.04
\,.
\label{212}
\end{align}
Therefore,
the energy splitting is extremely small.
The authors of Ref.~\cite{0801.2079}
noted that the splitting of the two hyperfine $1s$ energy levels
of the electron is many order of magnitude too large
(and the contribution to the decay of one of the two states is
suppressed by angular momentum conservation).
It is difficult to find a mechanism which produces a smaller
energy splitting.
Furthermore,
since the amplitude $A\simeq0.2$ of the interference is rather small,
it is necessary to find a mechanism which
generates coherently the states
$\mathbb{I}_{1}$ and $\mathbb{I}_{2}$
with probabilities
$|\mathcal{A}_{1}|^2$ and $|\mathcal{A}_{2}|^2$
having a ratio of about 1/99!

In conclusion, I have shown that
the standard method of calculation of the rates
(cross sections and decay rates)
of interaction processes
by summing over the rates of production
of all the allowed channels with a defined number of particles in the final state,
regardless of a possible coherence among them,
is correct.
The argument has been clarified through an analogy with
a double-slit experiment, emphasizing that it is a consequence of causality.
I have explained the reasons why the claim in Refs.~\cite{0801.1465,0801.2121,0801.3262}
that the GSI time anomaly is due to the mixing of neutrinos
in the final state of the electron-capture process
is incorrect.
I have also shown that
the GSI time anomaly may be due to
quantum beats due to the existence of two
coherent energy levels of the decaying ion.
However,
since the required energy splitting is extremely small
(about $10^{-15}\,\text{eV}$)
and the two energy levels must be produced with relative probabilities having a ratio of about $1/99$,
finding an appropriate mechanism is very difficult.

\section*{Acknowledgments}

I enjoyed illuminating and stimulating discussions with
A. Faessler,
J. Jochum,
M. Laveder,
Yu.A. Litvinov,
Z.K. Silagadze.
I would like to thank the Department of Theoretical Physics of the University of Torino
for hospitality and support.

\raggedright


\begin{thebibliography}{10}

\bibitem{0801.1465}
H.J. Lipkin,
\newblock (2008), \href{http://arxiv.org/abs/0801.1465}{\url{arXiv:0801.1465}}.

\bibitem{0801.2121}
A.N. Ivanov, R. Reda and P. Kienle,
\newblock (2008), \href{http://arxiv.org/abs/0801.2121}{\url{arXiv:0801.2121}}.

\bibitem{0801.3262}
M. Faber,
\newblock (2008), \href{http://arxiv.org/abs/0801.3262}{\url{arXiv:0801.3262}}.

\bibitem{0801.2079}
Y. Litvinov et~al.,
\newblock (2008), \href{http://arxiv.org/abs/0801.2079}{\url{arXiv:0801.2079}}.

\bibitem{Bilenky:1978nj}
S.M. Bilenky and B. Pontecorvo,
\newblock Phys. Rep. 41 (1978) 225.

\bibitem{Bilenky:1987ty}
S.M. Bilenky and S.T. Petcov,
\newblock Rev. Mod. Phys. 59 (1987) 671.

\bibitem{hep-ph/9812360}
S.M. Bilenky, C. Giunti and W. Grimus,
\newblock Prog. Part. Nucl. Phys. 43 (1999) 1,
  \href{http://arxiv.org/abs/hep-ph/9812360}{\url{arXiv:hep-ph/9812360}}.

\bibitem{hep-ph/0202058}
M.C. Gonzalez-Garcia and Y. Nir,
\newblock Rev. Mod. Phys. 75 (2003) 345,
  \href{http://arxiv.org/abs/hep-ph/0202058}{\url{arXiv:hep-ph/0202058}}.

\bibitem{hep-ph/0310238}
C. Giunti and M. Laveder,
\newblock (2003),
  \href{http://arxiv.org/abs/hep-ph/0310238}{\url{arXiv:hep-ph/0310238}},
\newblock In ``Developments in Quantum Physics -- 2004'', p. 197-254, edited by
  F. Columbus and V. Krasnoholovets, Nova Science Publishers, Inc.

\bibitem{hep-ph/0405172}
M. Maltoni et~al.,
\newblock New J. Phys. 6 (2004) 122,
  \href{http://arxiv.org/abs/hep-ph/0405172}{\url{arXiv:hep-ph/0405172}}.

\bibitem{hep-ph/0506083}
G.L. Fogli et~al.,
\newblock Prog. Part. Nucl. Phys. 57 (2006) 742,
  \href{http://arxiv.org/abs/hep-ph/0506083}{\url{arXiv:hep-ph/0506083}}.

\bibitem{hep-ph/0606054}
A. Strumia and F. Vissani,
\newblock (2006),
  \href{http://arxiv.org/abs/hep-ph/0606054}{\url{arXiv:hep-ph/0606054}}.

\bibitem{Giunti-Kim-2007}
C. Giunti and C.W. Kim,
\newblock {{Fundamentals of Neutrino Physics and Astrophysics}} (Oxford
  University Press, 2007).

\bibitem{Giunti:1992cb}
C. Giunti, C.W. Kim and U.W. Lee,
\newblock Phys. Rev. D45 (1992) 2414.

\bibitem{hep-ph/0102320}
S.M. Bilenky and C. Giunti,
\newblock Int. J. Mod. Phys. A16 (2001) 3931,
  \href{http://arxiv.org/abs/hep-ph/0102320}{\url{arXiv:hep-ph/0102320}}.

\bibitem{hep-ph/0306239}
W. Alberico and S. Bilenky,
\newblock Phys. Part. Nucl. 35 (2004) 297,
  \href{http://arxiv.org/abs/hep-ph/0306239}{\url{arXiv:hep-ph/0306239}}.

\bibitem{hep-ph/0402217}
C. Giunti,
\newblock (2004),
  \href{http://arxiv.org/abs/hep-ph/0402217}{\url{arXiv:hep-ph/0402217}}.

\bibitem{hep-ph/0608070}
C. Giunti,
\newblock J. Phys. G: Nucl. Part. Phys. 34 (2007) R93,
  \href{http://arxiv.org/abs/hep-ph/0608070}{\url{arXiv:hep-ph/0608070}}.

\bibitem{0801.4639}
C. Giunti,
\newblock (2008), \href{http://arxiv.org/abs/0801.4639}{\url{arXiv:0801.4639}}.

\bibitem{0804.1099}
H. Burkhardt et~al.,
\newblock (2008), \href{http://arxiv.org/abs/0804.1099}{\url{arXiv:0804.1099}}.

\bibitem{Shrock:1980vy}
R.E. Shrock,
\newblock Phys. Lett. B96 (1980) 159.

\bibitem{McKellar:1980cn}
B.H.J. McKellar,
\newblock Phys. Lett. B97 (1980) 93.

\bibitem{Kobzarev:1980nk}
I.Y. Kobzarev et~al.,
\newblock Sov. J. Nucl. Phys. 32 (1980) 823.

\bibitem{Shrock:1981ct}
R.E. Shrock,
\newblock Phys. Rev. D24 (1981) 1232.

\bibitem{Shrock:1981wq}
R.E. Shrock,
\newblock Phys. Rev. D24 (1981) 1275.

\bibitem{Carter-Huber-2000}
R. Carter and J. Huber,
\newblock Chem. Soc. Rev. 29 (2000) 305,
\newblock URL:
  \href{{http://www.rsc.org/ej/CS/2000/a900724e.pdf}}{\url{{http://www.rsc.org/ej/CS/2000/a900724e.pdf}}}.

\end{thebibliography}
\end{document}